# Weakened magnetic braking as the origin of anomalously rapid rotation in old field stars

Jennifer L. van Saders[1,2,3], Tugdual Ceillier[4], Travis S. Metcalfe[5], Victor Silva Aguirre[6], Marc H. Pinsonneault[7,3], Rafael A. García[4,3], Savita Mathur[5,3], Guy R. Davies[6,8]

[1] Carnegie-Princeton Fellow, Carnegie Observatories, 813 Santa Barbara Street, Pasadena, CA 91101

[2] Department of Astrophysical Sciences, Princeton University, Princeton, NJ 08544, USA

[3] Kavli Institute for Theoretical Physics, University of California, Santa Barbara, CA 93106-4030, USA

[4] Laboratoire AIM, CEA/DSM -- CNRS - Univ. Paris Diderot -- IRFU/SAp, Centre de Saclay, 91191 Gif-sur-Yvette Cedex, France

[5] Space Science Institute, 4750 Walnut Street Suite 205, Boulder CO 80301 USA

[6] Stellar Astrophysics Centre, Department of Physics and Astronomy, Aarhus University, Ny Munkegade 120, DK-8000 Aarhus, Denmark

[7] Ohio State University, Dept. of Astronomy, 140 W. 18th Ave. Columbus, OH 43210

[8] School of Physics and Astronomy, University of Birmingham, Birmingham, B15 2TT, United Kingdom.

**A knowledge of stellar ages is crucial for our understanding of many astrophysical phenomena, and yet ages can be difficult to determine. As they become older, stars lose mass and angular momentum, resulting in an observed slowdown in surface rotation[1]. The technique of 'gyrochronology' uses the rotation period of a star to calculate its age[2,3]. However, stars of known age must be used for calibration, and, until recently, the approach was untested for old stars (older than 1 gigayear, Gyr). Rotation periods are now known for stars in an open cluster of intermediate age[4] (NGC 6819; 2.5 Gyr old), and for old field stars whose ages have been determined with asteroseismology[5,6]. The data for the cluster agree with previous period–age relations[4], but these relations fail to describe the asteroseismic sample[7]. Here we report stellar evolutionary modelling[5, 6, 8, 9, 10], and confirm the presence of unexpectedly rapid rotation in stars that are more evolved than the Sun. We demonstrate that models that incorporate dramatically weakened magnetic braking for old stars can—unlike existing models—reproduce both the asteroseismic and the cluster data. Our findings might suggest a fundamental change in the nature of ageing stellar dynamos, with the Sun being close to the critical transition to much weaker magnetized**

**winds. This weakened braking limits the diagnostic power of gyrochronology for those stars that are more than halfway through their main-sequence lifetimes.**

There are two approaches to the calibration and testing of gyrochronology. The first is a purely empirical approach, which utilizes a sample of stars with independently measured ages and rotation periods to construct a period-age relationship. These relationships are generally simple power-laws in age, period, and some mass-dependent quantity, and have seen wide usage[1,2,4,5,7]. The second, model-based approach uses stellar models and a prescription for magnetic braking to account for the functional dependence on all relevant stellar quantities, but relies on calibrators to determine the magnitude of the angular momentum loss. For this reason, it is well-suited to calibrating samples that only sparsely cover parameter space. It furthermore provides a method to attach physical meaning to observed braking behavior.

Magnetic braking prescriptions are typically scaled from the solar case; the Skumanich relation[1] yields angular momentum loss of the form $dJ/dt \propto \omega^3$, where $\omega$ is the angular rotation velocity[11]. These relations often use the dimensionless Rossby number, defined as the ratio of the period to the convective overturn timescale, $Ro = P/\tau_{cz}$ to characterize departures from this simple power law. Rossby number thresholds and scalings are routinely invoked to parameterize the magnetic field strength[12,13], the mass and composition dependence to the spin-down[2,14], observed saturation of the magnetic braking in rapid rotators, and the sharp transition from slow to rapid rotation in hot stars (>6250 K) due to thinning convective envelopes[14]. Under traditional prescriptions, stars undergo braking throughout their main sequence lifetimes, regardless of rotation rate. Observations of young and intermediate age open clusters have indicated that such treatments are reasonable[4,15]. However, the combination of long period, low amplitude signatures of rotation, and the challenge of age measurements in field stars resulted in a dearth of old stars to test such relationships. *Kepler* data provides a first test of these prescriptions in stars older than the Sun.

The high precision, long-baseline light curves from *Kepler* make such investigations possible. The rotation of the star manifests itself in *Kepler* data as a periodic modulation in the intensity as dark starspots rotate into and out of view. Intensity variations due to stellar oscillations are likewise present in the lightcurve, on shorter timescales. Low degree modes of oscillation probe the conditions of the deep stellar interior and internal structure of the star, providing ages that are precise to better than 10% in stars where many oscillation modes are detected at high signal to noise[16].

The first efforts to calibrate the gyrochronology relations using seismic targets uncovered tension between the cluster and seismic samples[7]. Although the form of the mass-period-age relation used in this study was similar to those in previous studies[2,4], the range of ages and more sophisticated treatment of observational uncertainties made it possible to determine that the sample did not obey a single power-law period-age relation. However, even this approach has limitations: it does not account for metallicity, changes in the stellar moment of inertia, and relied on a sample in which some stars did not have detailed seismic modeling or spectroscopy, biasing the seismic ages.

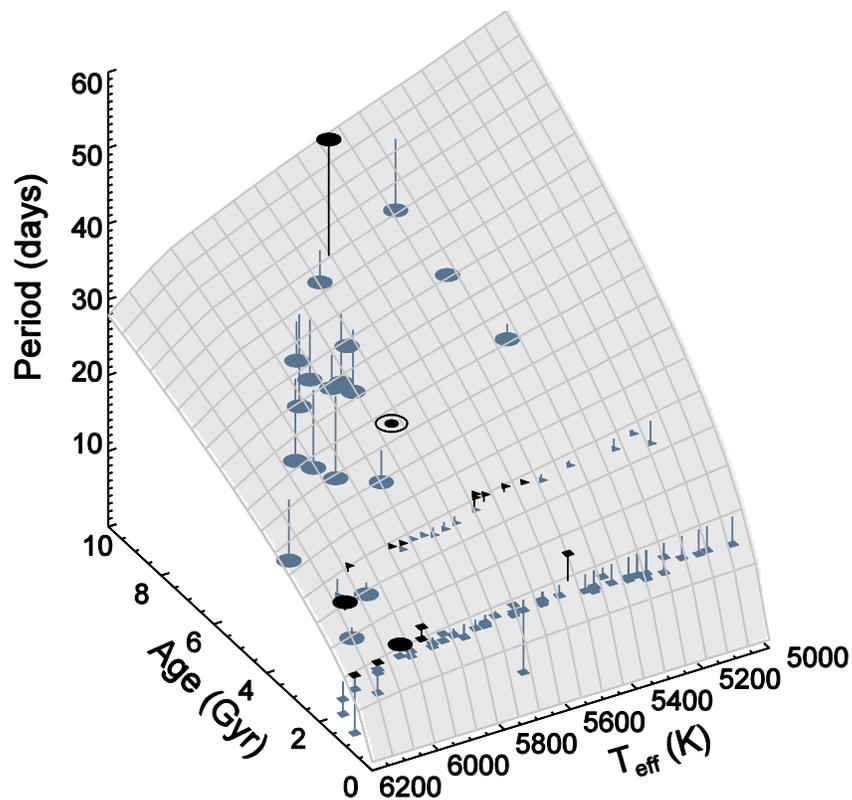

**Figure 1 | The gyrochronology period-age plane compared to observed periods.** The empirical gyrochronology relation[2,4] is shown as a plane. Open cluster data is shown as small squares (NGC6811, 1 Gyr) and triangles (NGC6819, 2.5 Gyr). Large circles represent the 21 star seismic sample, which falls systematically below the plane. The solar symbol marks the Sun, which falls on the plane by design.

To address the limitations of previous work and take full advantage of precisely determined stellar parameters, we utilize a subset of 21 *Kepler* stars selected to have detailed asteroseismic modeling and high precision ages, measured rotation periods, and measured metallicities[5-6,8-10], and couple these observations to stellar evolutionary models. The sample selection, details of the modeling to derive asteroseismic ages, and the extraction of the rotation periods are described in the Methods section. Figure 1 shows the surface in period, age, and effective temperature ($T_{eff}$, a proxy for mass) upon which stars are expected to lie[2,4]. Actual cluster and seismic data are overplotted, and while the clusters and young asteroseismic targets lie close to the plane, the intermediate age and old asteroseismic stars are strikingly discrepant and nearly all lie below the surface, rotating more rapidly than expected. When we account for uncertainties in the ages, masses, and compositions (see Methods) and predict the periods we should have observed given existing period-age relations[2,14] ($P_{expected}$), we find that the systematic offset persists: stars of roughly solar age and older are more rapidly rotating than predicted, regardless of the chosen period-age relation. Figure 2 highlights the systematic offset by plotting the ratios of the expected to observed periods for each star in the sample, where the expected periods are calculated using stellar models with a braking law calibrated on the Sun and open clusters[14] (a similar plot is provided in the Extended Data section for the empirical relation[2]). The theoretical models[14] fit the data with $\chi^2 = 54.9$, whereas the empirical relation[2] yields a $\chi^2 = 155.6$. In both cases, the systematic offset towards short rotation periods is an indication that the models predict more angular momentum loss than actually occurs. We therefore conclude that magnetic braking is weaker in these intermediate age and old stars. We extend our model by postulating that in addition to the Rossby scaling already present in the theoretical models[14], effective angular momentum loss ceases above a critical Rossby threshold[12]. We modify the angular momentum loss prescription[14] to conserve angular momentum above a specified $Ro_{crit}$. Visualizations of the effects of varying $Ro_{crit}$ values on the models are provided in Figure 3. The inclusion of the threshold has the desired effect: it reproduces the existing gyrochronology relations and cluster data at young ages, when Ro is smaller due to more rapid rotation, but allows stars to maintain unusually rapid rotation at late times. Furthermore, it reproduces the trend in mass (and zero-age main sequence (ZAMS) $T_{eff}$, which selects stars with similar rotational histories; we perform all fits using the seismic mass, but use ZAMS $T_{eff}$ for display to simplify the diagrams.) apparent in Figs. 2 and 3. Hotter, more massive stars reach the critical Rossby threshold at younger ages, and we therefore see discrepancies between the fiducial gyrochronology relationships and the observations at earlier times in panels of increasing ZAMS $T_{eff}$. The best-fit value for the Rossby threshold given our sample is $Ro_{crit} = 2.16 \pm 0.09$ ($\chi^2 = 13.3$) for the modified models. Shaded gray regions in both Figures 2 & 3 denote the full range of period ratios ($P_{Ro_{crit}}/P_{fiducial}$), and period-age combinations allowed for a model with $Ro_{crit} = 2.16$, given the ZAMS T$_{eff}$ range represented in each panel. These regions encompass all combinations of mass (0.4-2.0 M$_\odot$) and metallicity (-0.4 < [Z/H] < +0.4) that together produce a star within the appropriate ZAMS T$_{eff}$ range for each panel, on both the main sequence and subgiant branch.

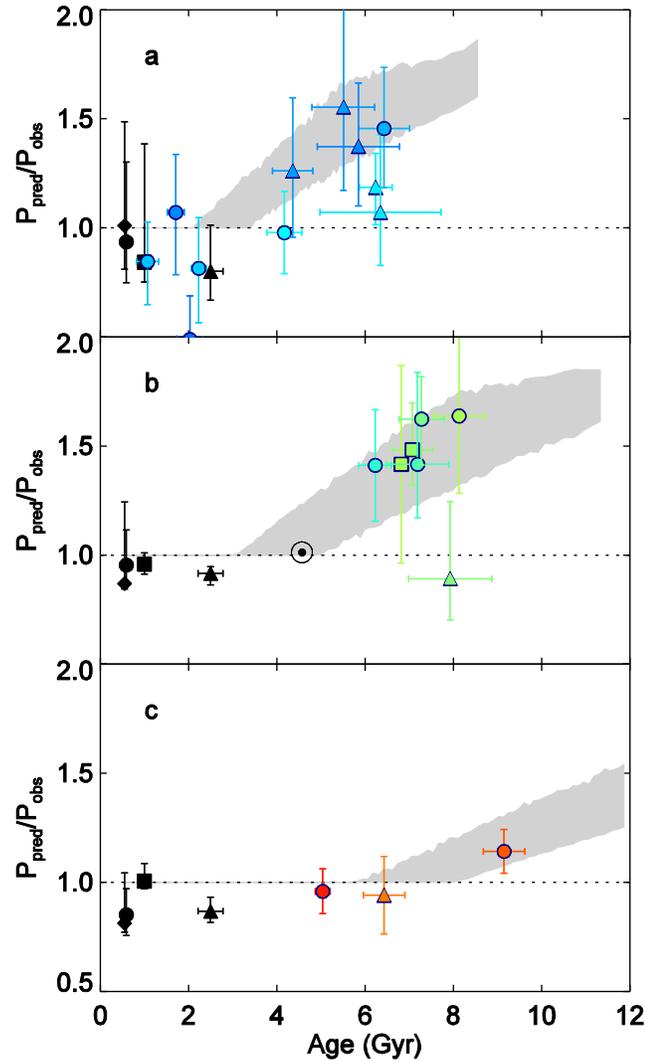

**Figure 2 | The ratio of the predicted rotation period[14] to the observed period.** Stars are divided into panels of decreasing AMP ZAMS T$_{eff}$ (a. 5900-6200 K, b. 5600-5900 K, c. 5100-5400 K). Period ratios for open clusters are shown as black symbols: M37 (diamond), Praesepe (circle), NGC6811 (square), NGC6819 (triangle). The Sun (☉) is marked. Colored circles represent seismic targets, colored triangles known planet hosts, and colored squares 16 Cyg A & B. All errors are 1σ. Stars are colored by ZAMS T$_{eff}$, with blue representing hotter stars. Shaded regions represent the period ratios permitted in each T$_{eff}$ bin for a Ro$_{crit}$ = 2.16 model.

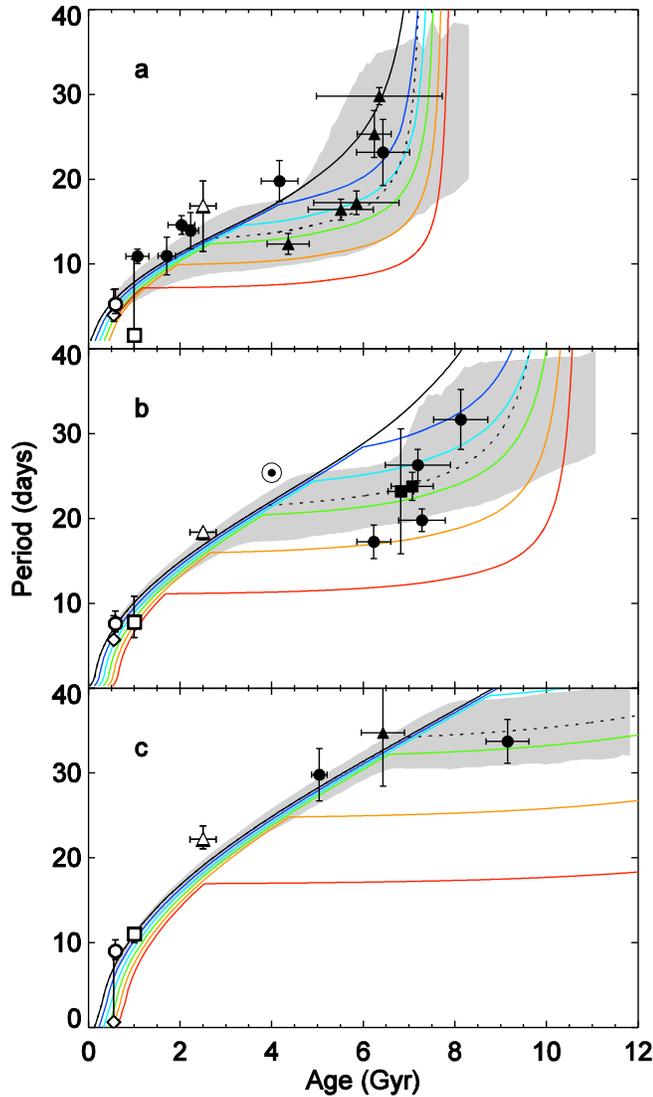

**Figure 3 |The effects of a Ro_crit threshold on rotational evolution.** Panel divisions and symbol conventions are adopted from Figure 2. Model curves are shown for solar metallicity and ZAMS $T_{eff}$ 6050K, 5750K, and 5250K, respectively. Curves are color-coded by $Ro_{crit}$: no $Ro_{crit}$ cut (black), 1.0 (dark blue), 1.5 (light blue), 2.0 (green), 2.5 (orange), 3.0 (red), 2.16 (dashed black). Successive curves are offset by +0.1 Gyrs to improve readability. Seismic (cluster) targets are overplotted in solid (open) symbols with 1σ errors. Shaded regions represent $Ro_{crit} = 2.16$ models for each $T_{eff}$ range.

We emphasize that our result—that old stars are too rapidly rotating—persists regardless of the choice of literature period-age relationship, asteroseismic modeling pipeline, or model uncertainties (see Methods). The period detection algorithms[17] and seismic ages have been well-tested[10]. The tight rotational sequences observed in intermediate age open clusters[4] suggest that we are not simply detecting the rapidly rotating tail of a population with a wide distribution of rotation rates, and it is unlikely that our stars with detected rotation are atypical (see Methods for further discussion).

Our model represents the limiting case in which the braking is so ineffective that the star ceases to shed angular momentum. If we instead allow the exponent, $\alpha$, of the period-age relation $P \propto t^{1/\alpha}$ to vary while fixing $Ro_{crit} = Ro_\odot = 2.16$, we do not obtain a comparable fit in the old stars until $\alpha \gtrsim 20$, suggesting that the braking is indeed drastically reduced. However, we observe spot modulation in these stars, which implies at least small-scale magnetic activity. The starspot properties may or may not directly reflect changes in the large scale field that governs spin-down. A change in field geometry from a simple dipole to higher order fields could produce weakened braking[18,19], as could a change in the distribution of spots on the stellar surface[20]. It could also be the case that the large-scale field strength undergoes a transition at high Rossby numbers[12]. Abrupt changes in the efficiency of angular momentum loss have been proposed in order to explain the rotational distributions in young clusters[21], and there is evidence for a Rossby-number-governed shift in the field morphologies in low-mass M dwarfs[22]. Observations of detailed magnetic field morphologies and corresponding simulations are lacking for stars at higher Rossby numbers than the Sun, and both are critical to understanding the source of the observed anomalous rotation.

Regardless of the mechanism that governs the spin-down, the observation that existing rotation-age relationships do not predict the observed rotation rates has immediate implications for gyrochronology. The rotation periods of the middle-aged stars that have passed this Rossby threshold represent only lower limits on the age. The empirical calibrations must be modified, and the weakened relationship between period and age will result in substantially more uncertain rotation-based ages for stars in the latter halves of their lives. The presence of such a Rossby threshold defines boundaries in mass-age space past which gyrochronology is incapable of delivering precise ages.

**Acknowledgements** We wish to thank the KITP and the organizers of the "Galactic Archaeology and Precision Stellar Astrophysics" program held from January to April 2015. We thank B. Shappee for useful discussion. T.S.M acknowledges the adopt-a-star crowdfunding program administered by White Dwarf Research Corp. G.R.D. acknowledges the support of the UK Science and Technology Facilities Council (STFC). M.H.P, T.S.M, and S.M. would like to acknowledge support from NASA grant NNX15AF13G and NSF grant AST-1411685. T.C. and R.A.G. received funding from the CNES grants at CEA. R.A.G. also acknowledges the funding from the European Community Seventh Framework Programme ([FP7/2007−2013]) under grant agreement No. 312844 (SPACEINN) and No. 269194 (IRSES/ASK). This research was supported in part by the National Science Foundation under Grant No. NSF PHY11-25915. Funding for the Stellar Astrophysics Centre is provided by The Danish National Research Foundation (Grant agreement no.: DNRF106). The research is supported by the ASTERISK project (ASTERoseismic Investigations with SONG and *Kepler*) funded by the European Research Council (Grant agreement no.: 267864). V.S.A. acknowledges support from VILLUM FONDEN (research grant 10118).

**Author Contributions** J.vS provided the physical interpretation of the rapid rotation, rotational and stellar modeling. M.H.P. contributed to the design of the rotation tracer code, ongoing YREC development, and interpretation of results. T.S.M. provided AMP modeling of all targets, and V.S.A. BASTA-GARSTEC modeling. T. C., R.A.G. and S.M. developed and implemented the analysis of rotational modulation in *Kepler* data. G.R.D provided updated asteroseismic mode frequencies used in the modeling.

# Methods

## Sample Selection

Our sample can be divided into two principal target types: *Kepler* Asteroseismic Science Consortium (KASC) targets, and *Kepler* Objects of Interest (KOI). We focus on those stars with (modeled) ZAMS $T_{eff}$s (defined as the point at which hydrogen fusion dominates the stellar luminosity) below 6200 K, where magnetic braking should be most important. We show the positions of the selected stars on a Hertzsprung Russell diagram in Extended Data Figure 1 and period-age plot in Extended Data Figure 2.

As described elsewhere[9], the asteroseismic sample is drawn from a magnitude-limited sample of 2000 solar-like stars that were selected for 1-month of short cadence (~1 minute) *Kepler* observations based on their properties in the *Kepler* Input Catalog (KIC). Of these stars, roughly 500 displayed evidence of solar-like oscillations. A subset of high signal-to-noise ratio targets was selected for continued monitoring over Q5-Q17. Of this sample, the mode frequencies for a subset of 61 high-signal-to-noise stars were extracted, and of these, 46 have high resolution spectroscopy. 42 of these stars were modeled with Asteroseismic Modeling Portal (AMP, described below), excluding 4 targets whose spectra contained a complicated pattern of mixed modes. 11 of these targets were both detected in spot modulation and were classified as "simple" solar-like oscillators that did not show the seismic hallmarks of F-stars and evolved subgiants. A further 3 (non-overlapping) targets were added[8]. Of this sample of 14, 12 targets have AMP ZAMS $T_{eff}$ < 6200 K, yielding a total of 12 stars in the KASC sample.

The KOI sample[10] was selected from the 77 KOIs observed in short cadence that displayed signatures of solar-like oscillations. Of these, 35 power spectra were of sufficient quality to extract individual mode frequencies to be modeled, 33 of which are unevolved main sequence stars. A subset of 11 have periods detected via spot modulation[6], 7 of which have an AMP ZAMS $T_{eff}$ < 6200 K.

Finally, we add the well-studied 16 Cyg A & B to our sample, which have asteroseismic ages[16] and rotation periods inferred from asteroseismic mode splittings[23]. In total, 21 stars are addressed in this analysis. Where available, we utilize the updated asteroseismic frequencies of ref. 24. We provide a table of the seismic (mass, age) and spectroscopic ($T_{eff}$, [Fe/H]) values and rotation periods in Extended Data Table 1.

**Age and Period Measurements**

Asteroseismic ages are determined using two methodologies. The Asteroseismic Modeling Portal (AMP), which provides the ages used in most of this paper, and BAyesian STellar Algorithm (BASTA) pipeline ages, used to verify that the discrepancies in predicted and observed rotation periods are not the result of pipeline choice. AMP uses a genetic algorithm to perform a search for the global χ2 minimum between the stellar observables and stellar model values[9]. The code utilizes the Aarhus stellar evolution code (ASTEC) and adiabatic pulsation code (ADIPLS) to compute oscillation frequencies. The BASTA pipeline uses a Bayesian approach to model stars with a grid of models produced with the Garching Stellar Evolution Code (GARSTEC). The input physics of the stellar models utilized in each method are detailed in their respective papers[8-10].

Both methods use frequency spacings and spectroscopic constraints to identify the optimal stellar properties, but AMP also uses the individual frequencies by employing an empirical correction for surface effects. There are two main differences between the models used by BASTA and those used by AMP. BASTA-GARSTEC uses a fixed relationship between the initial helium and metallicity, anchored to zero metallicity at the primordial helium abundance and assuming $\Delta Y/\Delta Z = 1.4$ to reproduce the solar values. It also uses a single solar-calibrated value of the mixing-length parameter for all models. AMP-ASTEC allows the initial helium to float independently of metallicity, and searches a wide range of values for the mixing-length parameter. Both sets of models include diffusion, although BASTA-GARSTEC includes both helium and heavy-metal diffusion, while AMP-ASTEC considers only helium diffusion.

We extract rotation periods using techniques[5] that we summarize briefly here (full period extraction diagrams are available at http://irfu.cea.fr/Phocea/Vie_des_labos/Ast/ast_technique.php?id_ast=3607). For the corrected lightcurve of each *Kepler* star, the autocorrelation function (ACF) and a wavelets decomposition (period-time) are calculated. We collapse the wavelet decomposition on the period axis to obtain the global wavelet power spectrum (GWPS), and the significant peaks of this GWPS are fitted with Gaussian functions. In parallel, we identify the most significant peaks of the ACF. The derived surface rotation period is the result of the comparison of the ACF and GWPS analyses and is confirmed by a visual inspection of the lightcurves.

## Stellar Rotation Models

We use a theoretical model grid[14] (using OPAL rather than OP opacities; all other inputs are unchanged), and utilize the same loss-law calibration and form, and assume solid body rotation. The model grid is expanded to cover a wider range of metallicities and masses, namely [Z/H] = -0.4 to [Z/H] = +0.4, assuming a helium enrichment of $\Delta Y/\Delta Z = 1.0$ and no diffusion or gravitational settling. We use the "fast launch" conditions[14] for modeling the rotation, but have validated that our results are insensitive to the choice of initial conditions. Changing the launch conditions typically shifts the period ratio ( in sense of expected/observed) by less than 50% of the quoted errors, and shifts the fitted critical Rossby number to $Ro_{crit} = 2.15 \pm 0.08$. The model $\tau_{cz}$ is the local convective overturn timescale, defined as the ratio of the typical mixing length to the convective velocity at one pressure scale height above the base of the convective envelope in the mixing length theory of convection. Under this definition, with $P_\odot = 25.4$ days, $\tau_{cz,\odot} = 1.015 \times 10^6$ s, $Ro_\odot = 2.16$.

The weakened magnetic braking is modeled by modifying the braking law such that a star with $P/\tau_{cz} > Ro_{crit}$ is evolved under the assumption of conservation of angular momentum, such that the rotation period depends only on the changing moment of inertia of the star as it evolves. The modified loss law is given by ref. 14 (following eqns. 1 & 2 in the reference):

$$\frac{dJ}{dt} = \begin{cases} f_K K_M \omega \left(\frac{\omega_{crit}}{\omega_\odot}\right)^2, \omega_{crit} \leq \omega \frac{\tau_{cz}}{\tau_\odot}, Ro \leq Ro_{crit} \\ f_K K_M \omega \left(\frac{\omega \tau_{cz}}{\omega_s \tau_{cz,\odot}}\right)^2, \omega_{crit} > \omega \frac{\tau_{cz}}{\tau_\odot}, Ro \leq Ro_{crit} \\ 0, Ro > Ro_{crit} \end{cases}$$

$$\frac{K_M}{K_{M,\odot}} = c(\omega) \left(\frac{R}{R_\odot}\right)^{3.1} \left(\frac{M}{M_\odot}\right)^{-0.22} \left(\frac{L}{L_\odot}\right)^{0.56} \left(\frac{P_{phot}}{P_{phot,\odot}}\right)^{0.44}$$

where $f_K$ is a constant factor to used to scale the loss law during the empirical fitting, $\omega_{crit}$ is the saturation threshold (important only at young ages), $\tau_{cz}$ is the convective overturn timescale, and $P_{phot}$ the pressure at the photosphere. The term $c(\omega)$ sets the centrifugal correction; because our stars are slowly rotating and the correction should be small, we set $c(\omega) = \text{const} = 1$. This braking law is fit to open cluster data and the Sun, where the initial rotation period, disk-locking timescale, $\omega_{crit}$, and $f_K$ were allowed to vary, and all other parameters were determined using stellar evolutionary models[14]. When fitting for an optimal $Ro_{crit}$, we keep the parameters of the magnetic braking law calibrated on the Sun and open clusters, and vary only the $Ro_{crit}$ at which braking is allowed to cease. $Ro_{crit}$ is optimized using a $\chi^2$ figure of merit (valid under the assumption of independent observations and Gaussian uncertainties): $\chi^2 = \sum_i^N (P_{obs,i} - P_{mod,i})^2/(\sigma_{obs,i}^2 + \sigma_{mod,i}^2)$, where $\sigma_{obs,i}$ is the observational uncertainty on the extracted period, and $\sigma_{mod,i}$ represents the uncertainty on the model period given the uncertainties on the input masses, ages, and compositions. We derive uncertainties on $Ro_{crit}$ using bootstrap resampling, drawing a 21 star sample with replacement from the original data 50000 times, and recalculating the best-fit $Ro_{crit}$ for each realization. Cluster data and the Sun are *not* utilized in this fit. An alternate fit allowing parameters important for late-time braking to vary ($f_K$, $Ro_{crit}$) and including intermediate age and older rotation data from the seismic sample, NGC6819, and the Sun, (52 stars in total) yields a best-fit $Ro_{crit} = 2.1\pm0.1$, with $f_K = 8.4\pm0.2$.

Predicted model periods are obtained by using the mass and age from the asteroseismic pipelines coupled with the spectroscopic metallicity[8-10,16]. Model uncertainties are estimated by generating 50000 (20000 for $Ro_{crit} + f_K$ fit) realizations of the input parameters ($M$, $t$, and [Fe/H]), where values are drawn from a Gaussian distribution centered on the observed value with $1\sigma$ errors defined by the observational uncertainties. While we search in the fundamental space of mass, age and composition, we only select models which fall within $5\sigma$ of the observed $T_{eff}$. This constraint has little or no effect for unevolved stars, but ensures that stars at the turnoff (KIC 6196457 and 8349582 in particular) are not assigned artificially long rotation periods due to mass-age combinations that fall on the subgiant branch. $1\sigma$ uncertainties on the model periods are defined as the values that enclose 68% of the resulting models.

### Empirical Gyrochronology Relations

We verify that the unexpectedly rapid rotation in old, solar-like stars is independent of the spin-down prescription by repeating our exercise with an empirical literature gyrochronology relation[2]. We replicate Figure 2 in the main body in Extended Data Figure 3 with predicted

periods drawn from an empirical gyrochronology calibration, which takes the form (ref. 2, eqn. 32):

$$t = \frac{\tau}{k_C} \ln\left(\frac{P}{P_0}\right) + \frac{k_I}{2\tau}(P^2 - P_0^2),$$

where $t$ is the age, $\tau$ is the convective overturn timescale, $P$ the period, and $P_0$ the initial period. We adopt values for the constants $k_C = 0.646$ Myr d$^{-1}$ and $k_I = 452$ d Myr$^{-1}$ and $P_0 = 1.1$ days[2,4], and the global $\tau$- $T_{eff}$ relation utilized in both works. 50000 realizations of the combination ($T_{eff}$, t) are drawn from a Gaussian distribution centered on the measured values, with a 1σ width defined by the quoted observational errors on the central values. These empirical relationships do not account for physical expansion of stars as they evolve (particularly near the end of the main sequence) and therefore tend to predict somewhat more rapid rotation than full theoretical models near the main sequence turnoff.

## Cluster Data

To provide comparison to the typical gyrochronological calibrators, we draw cluster data from a variety of literature sources. For the cluster M37 we adopt the cluster parameters[15] E(B-V) = 0.227±0.038 mag, [M/H] = 0.045±0.044 dex, an age of 550±30 Myr. Rotation data and cluster parameters[25] for Praesepe (M44) are included, with E(B-V) = 0.027±0.004, [Fe/H] = 0.11±0.03, and log(age) = 8.77±0.1. We adopt the *g-r* colors, E(B-V) = 0.1, and periods for NGC6811[26], and the cluster metallicity [M/H] = -0.1±0.01 and age, 1.00±0.05 Gyr[27]. Finally, we utilize NGC6819 periods and *B-V* colors[4], with the age (2.5±0.2 Gyr) and adopted metallicity (0.09±0.03)[28]. *B-V* colors are converted into temperatures and stellar masses using YREC isochrones[29]. We model cluster stars in the same manner as the seismic targets, with 10000 mass-age-composition realizations for each star. We display the mean cluster rotation periods for all stars within the ZAMS $T_{eff}$ bins, with errors representing the 16$^{th}$ and 84$^{th}$ percentiles. In M37 and Praesepe in particular, the rotational distribution displays a range resulting from spread in the initial rotation periods.

## Sample Biases

We demonstrate that our results are unlikely to be a consequence of selection bias in our sample. The sample is subject to two sources of selection bias: asteroseismic detectability, and the detectability of spot modulation.

Detailed asteroseismic analysis requires a high signal-to-noise detection of the power excess from oscillations. Oscillation amplitudes scale[30] roughly as $A_{max} \propto (L/M) (T_{eff})^{-2}$ (eqn. 7, referring to l = 0 radial modes); seismic samples are therefore strongly biased towards more massive stars. There is also a bias toward bright targets, where lower noise levels contribute to detectability. Our sample is drawn from two subsets of stars: the 1-month survey stars from the seismic sample, and the KOIs. We expect the 1-month survey seismic detections at magnitudes $K_p \lesssim 10$, while the roughly 1000 day timeseries in short cadence collected for the KOI sample allow detections out to $K_p \approx 12$, which well describes the actual magnitude distribution of our sample (see Fig. 6 in ref. 30). The strong trends with magnitude and mass are well-predicted by basic scaling arguments[30], save for the dependence on activity: active stars are *less* likely to be detected in oscillations[30]. Our sample is selected seismically, and we do not expect the well-understood seismic biases to favor rapid rotators (apart from the obvious mass dependence).

Variability due to starspots scales with the rotation period, in the sense that more rapid rotation is associated with higher amplitudes of variability[15]. One could imagine that we are detecting the rapidly rotating tail of a distribution of rotation periods, or detecting objects spun up by binary/planetary interactions or mergers.

This first case is at odds with what we know from open clusters: as late as 2.5 Gyr, there is a converged, well-defined rotational sequence that shows very little scatter at fixed mass[4]. If we are in fact detecting a rapidly rotating subset of the population, the dispersion in rotation and spin-down rates must set in after several Gyrs, or it would be visible in the open cluster data. If there is dispersion in the rotation periods, it represents a serious challenge to the validity of gyrochronology for old stars, regardless of its source.

The pipeline used to extract the rotation periods for this work has been tested with an injection and recovery exercise[17]. Our recovery fraction is shown in Extended Data Figure 4, and demonstrates that we should be able to detect stars that are substantially less active than the Sun at longer periods. However, this exercise does not account for stars that simply cease to have spots to detect on their surfaces; under this scenario, slow rotators could exist but be undetectable. We cannot directly combat this concern given our current dataset, although we can examine the case of 16 Cyg. 16 Cyg A & B are not detected in spot modulation; their periods are derived from asteroseismic frequency splittings, which yield periods that probe the envelope rotation[23]. If we assume that these stars have solar-like rotation profiles, the seismic rotation periods are directly comparable to the surface periods. This pair displays the same anomalously rapid rotation as objects detected in spot modulation, providing evidence against the argument that stars undetected in spot modulation are simply more slowly rotating. It is also worth noting that our own Sun would be undetectable during the minimum of its activity cycle (see CEA results[17]). Our non-detections could equally be the result of the normal variations in the activity of Sun-like stars, rather than a period bias.

Finally, we examine the possibility of interactions or mergers with other bodies. In our sample 16 Cyg A & B, KIC 3427720 and KIC 9139151 are known or suspected binaries[5]. In each case the components are well separated, the binary orbits are estimated well in excess of 10,000 years. In order for a companion to significantly affect the rotation, it must be at orbital periods comparable to the rotation period, and will therefore be unresolved. The KOI sample has undergone the extensive vetting associated with planet detection; all planets are confirmed, and there is no evidence of transit timing variations that would accompany a close stellar companion. System stability is unlikely for binary orbits of order 30 days that contain even a low mass stellar companion[31]. Likewise, there is no evidence for interaction between the planets and the host stars in the KOI sample[6], and no known Hot Jupiters. In the case of the seismic sample, there is no evidence for double lined binaries, photometric-spectroscopic temperature disagreements, multiple oscillating components or unusual dilution of the seismic power spectra, and no evidence of eclipses. Finally, if mergers (planetary or stellar) were responsible for all detections of rapid rotation, then the 50% detection rate of the "simple stars[9]" in spot modulation implies an uncomfortably high merger rate.

### The Asteroseismic Age Scale

We perform two tests to demonstrate that the discrepancy between the expected and observed rotation periods is not a systematic with roots in the asteroseismic age scale. We show that ages derived with the BASTA pipeline display the same trend in rotation period, and that systematically shifting the asteroseismic ages, while improving the fit, is inferior to instituting a Rossby threshold.

Figure 5 in the Extended Data provides period ratio plots using the BASTA ages and BASTA ZAMS $T_{eff}$ determinations. The systematic trend in the period ratios survives. The Barnes relation[2] fits with $\chi^2 = 184.3$, and the fiducial models[14] with $\chi^2 = 68.4$. A fit for $Ro_{crit}$ using the BASTA ages yields $Ro_{crit} = 2.67 \pm 0.50$. Bootstrap resampling demonstrates that this number is sensitive to whether KIC 8349582 is drawn; if KIC 8349582 is excluded, the fit becomes $Ro_{crit} = 2.12 \pm 0.12$.

We also investigate the possibility that the seismic age scale is systematically shifted relative to the true ages. We perform model fits with the fiducial braking law with an extra parameter that allows for a systematic age shift. For the AMP ages, $\chi^2$ is minimized with the Barnes relation with a systematic shift of 35% and $\chi^2 = 78.5$. Likewise, the fiducial models[14] prefer a shift of 20±3% with a $\chi^2 = 26.9$. In both cases, the required systematic shifts are larger than the estimated 9.6% systematic uncertainties in seismic ages[10].

Finally, to verify that we are not biased by the fact that the ages and periods were determined using different evolution codes, we tune the physics in the fiducial models[14] to match that of the AMP models, and predict the rotation periods for the central AMP values of the masses, ages, and compositions of each star. In particular, we match the diffusion physics, opacity tables, equation of state, helium and metal abundances, boundary conditions, and important nuclear reaction rates present in the ASTEC code used for AMP. The results are presented in Extended Data Figure 6, and demonstrate that the discrepancy between the predicted and observed periods is preserved. We conclude that our result is not the consequence of assumptions about the stellar physics included in models.

## Code Availability

The AMP science code used to infer stellar ages can be downloaded at https://amp.phys.au.dk/about/evolpack. Code for the period extraction and rotational evolution will be publically released upon completion of the necessary documentation. YREC likewise has no public documentation, and has not been publically released. BASTA is undergoing major revisions for increased speed and is not yet publically available.

## References con't

**Extended Data Table 1 | Rotation periods, asteroseismic, and spectroscopic quantities for sample stars.** Units are as follows: mass (M$_\odot$), age (Gyr), log(g) (g/cm$^2$), T$_{eff}$ (K), period (days). Quoted errors are 1σ.

| KIC | AMP Mass | AMP Age | AMP log(g) | AMP ZAMS T$_{eff}$ | BASTA/GARSTEC Mass | BASTA/GARSTEC Age | BASTA/GARSTEC ZAMS T$_{eff}$ | Spectroscopic T$_{eff}$ | Spectroscopic [Fe/H] | Period | Note |
|---|---|---|---|---|---|---|---|---|---|---|---|
| 16Cyg A | 1.10±0.02 | 7.07±0.46 | 4.295 | 5677 | 1.04±0.01 | 6.95±0.26 | 5668 | 5825±50 | +0.09±0.02 | 23.8±1.7 | seismic |
| 16Cyg B | 1.06±0.02 | 6.82±0.28 | 4.360 | 5629 | 0.998±0.005 | 7.02±0.14 | 5592 | 5750±50 | +0.05±0.02 | 23.2±7.4 | seismic |
| 3427720 | 1.13±0.04 | 2.23±0.17 | 4.388 | 5985 | 1.12±0.02 | 2.22±0.31 | 6019 | 6040±84 | -0.03±0.09 | 13.9±2.1 | seismic |
| 3656476 | 1.17±0.03 | 8.13±0.59 | 4.246 | 5642 | 1.07±0.01 | 7.68±0.42 | 5525 | 5710±84 | +0.25±0.09 | 31.7±3.5 | seismic |
| 5184732 | 1.27±0.04 | 4.17±0.40 | 4.270 | 5905 | 1.18±0.02 | 4.05±0.42 | 5810 | 5840±84 | +0.38±0.09 | 19.8±2.4 | seismic |
| 6116048 | 1.01±0.03 | 6.23±0.37 | 4.270 | 5838 | 1.06±0.02 | 5.54±0.34 | 5943 | 5935±84 | -0.24±0.09 | 17.3±2.0 | seismic |
| 6196457 | 1.23±0.04 | 5.51±0.71 | 4.053 | 6064 | 1.21±0.02 | 5.52±0.50 | 5991 | 5871±94 | +0.17±0.11 | 16.4±1.2 | KOI |
| 6521045 | 1.04±0.02 | 6.24±0.37 | 4.118 | 5933 | 1.11±0.02 | 6.50±0.51 | 5886 | 5825±75 | +0.02±0.10 | 25.3±2.8 | KOI |
| 7680114 | 1.13±0.03 | 7.19±0.70 | 4.184 | 5801 | -- | -- | -- | 5855±84 | +0.11±0.09 | 26.3±1.9 | seismic |
| 7871531 | 0.84±0.02 | 9.15±0.47 | 4.479 | 5253 | 0.84±0.02 | 10.10±0.99 | 5240 | 5400±84 | -0.24±0.09 | 33.7±2.6 | seismic |
| 8006161 | 1.04±0.02 | 5.04±0.17 | 4.502 | 5165 | 0.948±0.005 | 5.08±0.10 | 5250 | 5390±84 | +0.34±0.09 | 29.8±3.1 | seismic |
| 8349582 | 1.19±0.04 | 7.93±0.94 | 4.178 | 5695 | 1.07±0.02 | 8.03±0.75 | 5630 | 5699±74 | +0.30±0.10 | 51.0±1.5 | KOI |
| 9098294 | 1.00±0.03 | 7.28±0.51 | 4.314 | 5718 | 1.01±0.02 | 6.93±0.57 | 5734 | 5840±84 | -0.13±0.09 | 19.8±1.3 | seismic |
| 9139151 | 1.14±0.03 | 1.71±0.19 | 4.376 | 6092 | 1.16±0.02 | 1.79±0.46 | 6019 | 6125±84 | +0.11±0.09 | 11.0±2.2 | seismic |
| 9955598 | 0.96±0.01 | 6.43±0.47 | 4.506 | 5307 | 0.89±0.01 | 6.98±0.45 | 5250 | 5460±75 | +0.08±0.10 | 34.7±6.3 | KOI |
| 10454113 | 1.19±0.04 | 2.03±0.29 | 4.315 | 6138 | 1.15±0.03 | 2.86±0.54 | 6095 | 6120±84 | -0.06±0.09 | 14.6±1.1 | seismic |
| 10586004 | 1.16±0.05 | 6.35±1.37 | 4.072 | 5943 | 1.18±0.03 | 6.43±0.62 | 5753 | 5770±83 | +0.29±0.10 | 29.8±1.0 | KOI |
| 10644253 | 1.13±0.05 | 1.07±0.25 | 4.402 | 6001 | 1.16±0.02 | 1.20±0.39 | 5991 | 6030±84 | +0.12±0.09 | 10.91±0.87 | seismic |
| 10963065 | 1.07±0.02 | 4.36±0.46 | 4.294 | 6063 | 1.09±0.02 | 4.18±0.44 | 6076 | 6104±74 | -0.20±0.10 | 12.4±1.2 | KOI |
| 11244118 | 1.10±0.05 | 6.43±0.58 | 4.077 | 6023 | 1.13±0.02 | 6.90±0.44 | 5677 | 5745±84 | +0.35±0.09 | 23.2±3.9 | seismic |
| 11401755 | 1.03±0.05 | 5.85±0.93 | 4.043 | 6094 | 1.06±0.03 | 7.10±0.60 | 6057 | 5911±66 | -0.20±0.06 | 17.2±1.4 | KOI |

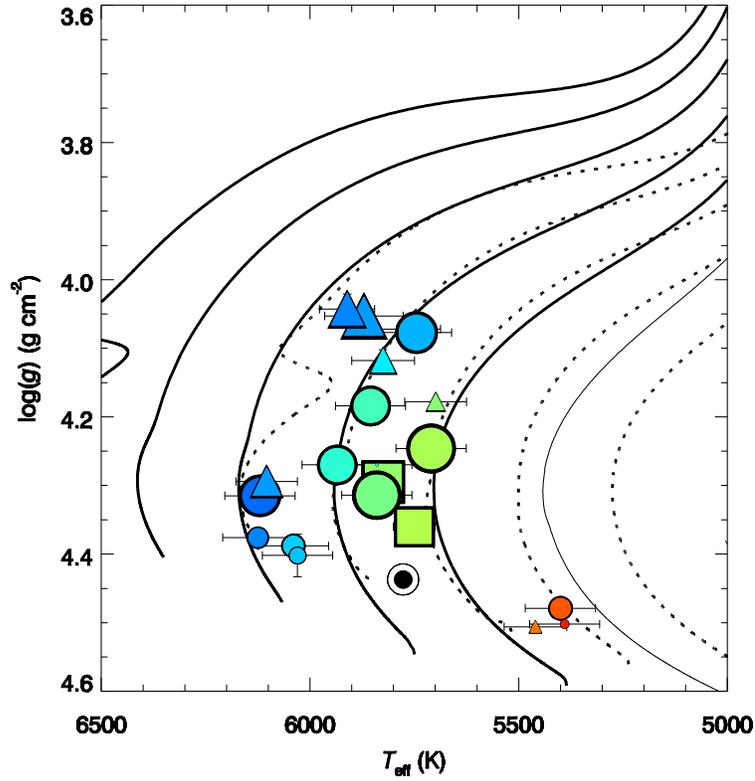

**Extended Data Figure 1 | The positions of all 21 stars on the Hertzsprung-Russell diagram.** We plot spectroscopic $T_{eff}$ versus seismic log(g) with 1σ observational errorbars, with the symbol size is proportional to the period ratio (AMP ages, fiducial models[14]). Points are color-coded in the same manner as Figure 2 in the body of the letter, and symbol conventions are retained. Evolutionary tracks for [Z/H] = +0.3 (dotted) and -0.1 (solid) for masses 0.8-1.3 $M_\odot$ in increments of 0.1 $M_\odot$ are overplotted ([Z/H] = +0.3, M = 0.8 $M_\odot$ is beyond the plot area).

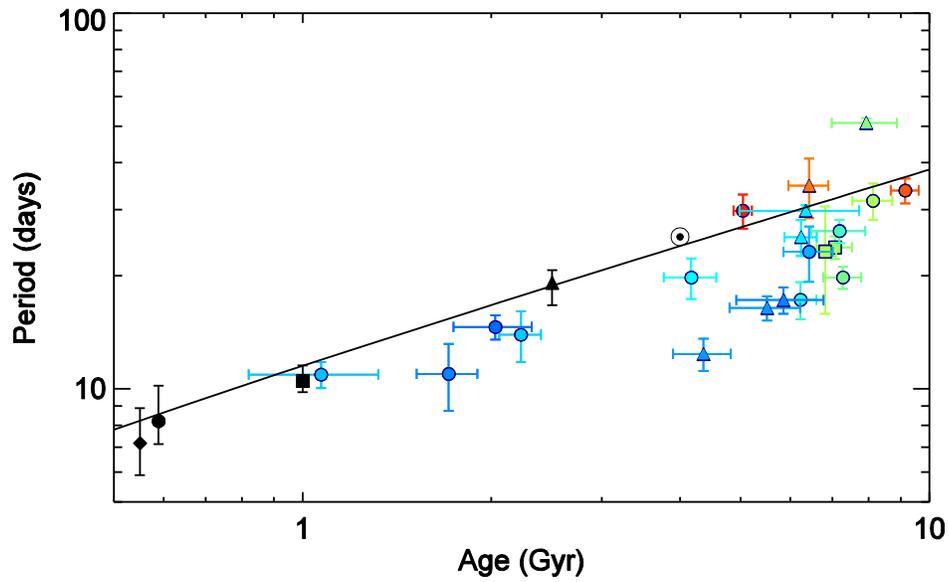

**Extended Data Figure 2 | Period-age plot of sample stars.** The 21 star sample shown in observed period and AMP asteroseismic ages. Symbol conventions are identical to those in the main body. The solid line denotes the empirical relation[2] for $T_{eff}$ = 5800 K (approximately equal to the mean sample $T_{eff}$). All errors are 1$\sigma$.

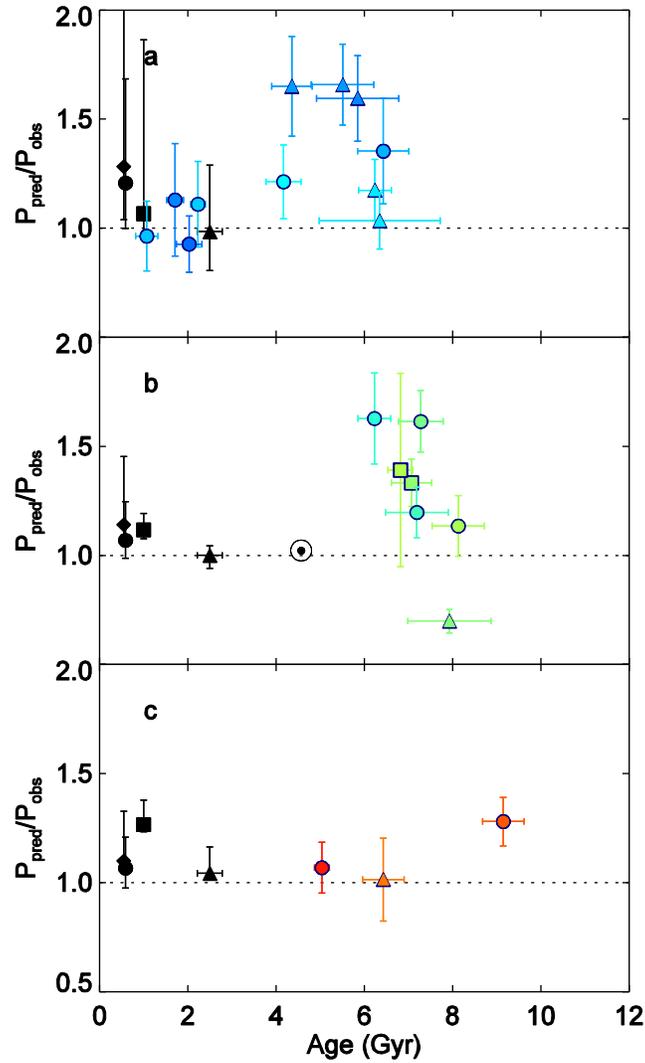

**Extended Data Figure 3 | Period ratios using empirical gyrochronology relations.** The ratio of the predicted periods[2] and the observed periods, as a function of the AMP asteroseismic age and AMP ZAMS $T_{eff}$. All errors are 1σ. The symbol conventions are identical to those in Fig. 2 in the body of the letter. Panel a displays ZAMS $T_{eff}$ 5900-6200 K, b. 5600-5900 K, and c. 5100-5400 K.

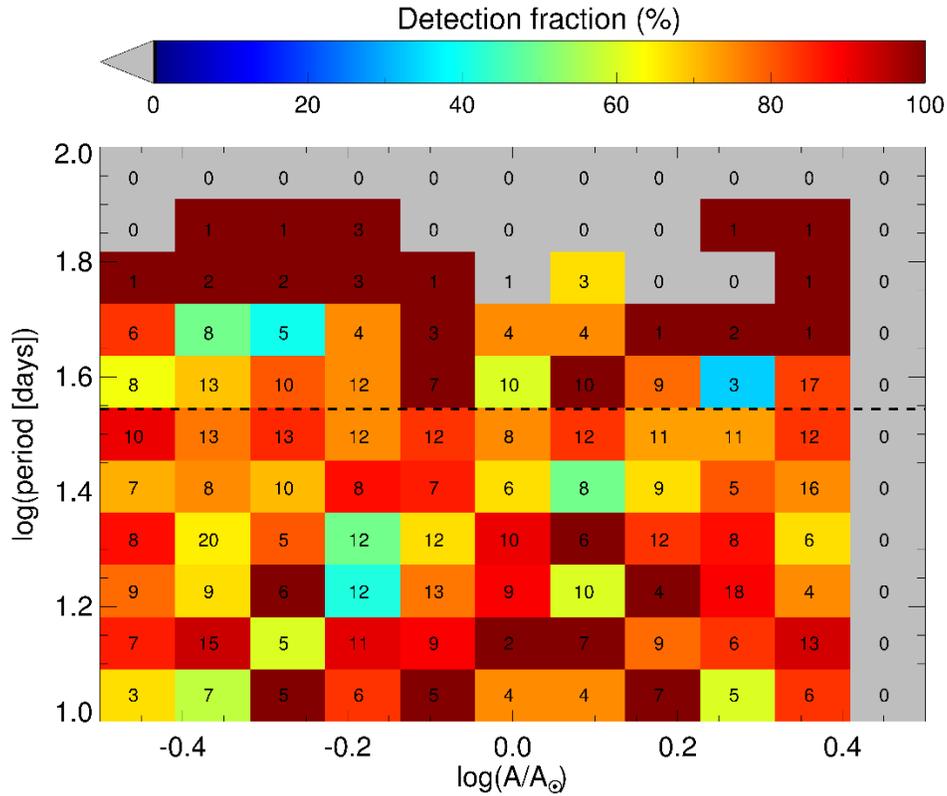

**Extended Data Figure 4 | Detectability of stars in spot modulation.** Detection fractions for the 750 stars with noise in the hound-and-hare exercise[17] as a function of activity level A (defined as $A_\odot = 1$) and period. The total number of lightcurves searched for periodicity in each cell is overplotted. The dashed black line at P = 35 days represents the expected period for stars like the Sun under traditional literature gyrochronology relations.

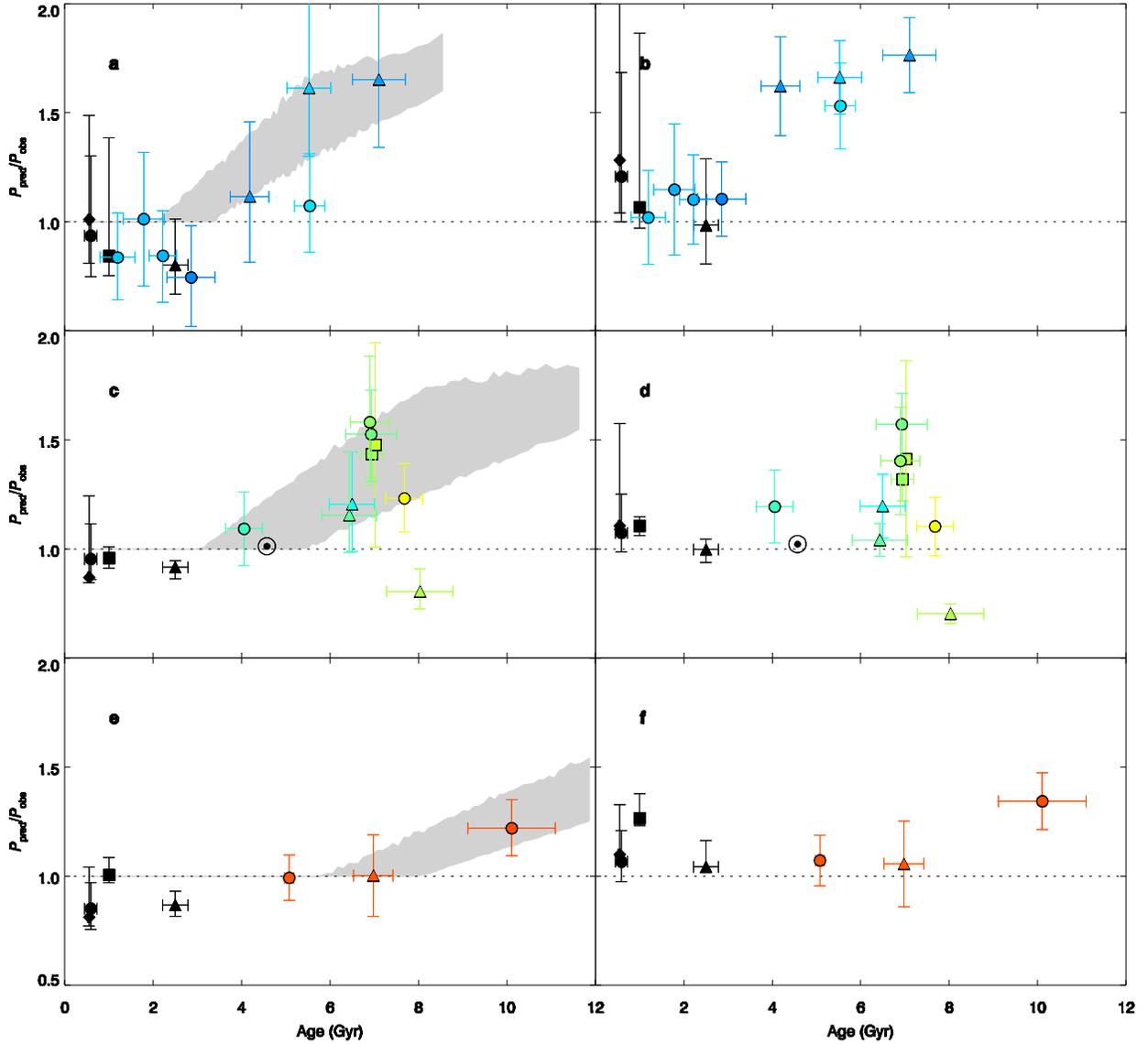

**Extended Data Figure 5 | Predicted versus observed periods using ages determined with BASTA.** The left-hand (a, c, e) column displays the comparison with the fiducial models[14], as well as the gray band representing the offset expected with Ro$_{crit}$ = 2.16 models. All errors are 1σ. The right-hand panels (b, d, f) compare the predicted periods from the empirical relation[2] to the observed periods. Stars are divided by ZAMS T$_{eff}$, using BASTA ZAMS T$_{eff}$ values. All symbol conventions and panel divisions are the same as those of Fig. 2, except the temperature range represented in the middle panels is now 5500K-5900K.

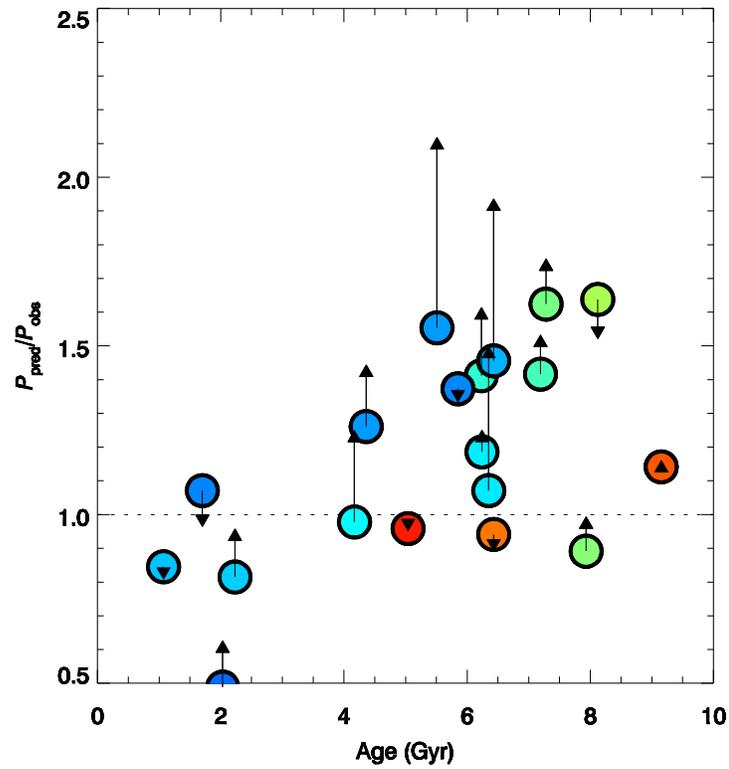

**Extended Data Figure 6 | The shift in the period ratios due to the differences in stellar model input physics.** Circles are color-coded by ZAMS $T_{eff}$, and represent the period ratio of the fiducial model[14] and observed periods. Arrows denote the shift in the period ratio when YREC models[14] are run matching the AMP-ASTEC physics.